\documentclass[twocolumn,showpacs,preprintnumbers,amsmath,amssymb]{revtex4}
\usepackage{dcolumn}
\usepackage{bm}
\usepackage[dvips]{graphicx}
\usepackage{wrapfig,subfigure}
\usepackage{amssymb}
\usepackage{color}
\def\tr{t_{\rm r}}

\def\W{W}
\def\ddq{\delta \dot q}
\def\st{\sigma^2}

\newcommand{\tc}{t_{\rm corr}}

\begin{document}

\title{Exponential peak and scaling of work fluctuations in modulated systems}
\author{M.I. Dykman}
\affiliation{
Department of Physics and Astronomy, Michigan State
 University, East Lansing, MI 48824, USA}
\date{\today}

\begin{abstract}
We extend the stationary-state work fluctuation theorem to
periodically modulated nonlinear systems. Such systems often have
coexisting stable periodic states. We show that work fluctuations
sharply increase near a kinetic phase transition where the state
populations are close to each other. The work variance
is proportional here to the reciprocal rate of interstate switching.
We also show that the variance displays scaling with the distance to
a bifurcation point and find the critical exponent for a saddle-node
bifurcation.
\end{abstract}

\pacs{05.40.-a, 05.70.Ln, 74.50.+r, 85.85.+j}

\maketitle


Since their discovery in the early 90s
\cite{Evans1993,Evans1994,Gallavotti1995a}, fluctuation theorems
have been attracting increasing interest. They establish general
features of fluctuating systems away from thermal equilibrium, see
Refs.~\onlinecite{Evans2002,Bustamante2005} for reviews. A major
``test bed" for fluctuation theorems is provided by dynamical
systems with a few degrees of freedom coupled to a thermal bath, a
Brownian particle being an example. Much of the corresponding
theoretical and experimental work refers to (i) modulated linear
systems, where fluctuations have been studied both in transient and
stationary regimes
\cite{Wang2002,Zon2004b,Garnier2005,Zamponi2005a,Cohen2007,Joubaud2007a,Imparato2007},
and (ii) nonlinear systems, initially at thermal equilibrium, driven
to a different, generally nonequilibrium state
\cite{Jarzynski1997,Crooks1999,Hummer2001,Collin2005,Blickle2006}.

Fluctuations in nonequilibrium dynamical systems have been
attracting attention also in a different context. They play an
important role in various types of mesoscopic vibrational systems of
current interest. Because damping of the vibrations is typically
weak, even a moderately strong resonant force can excite them to
comparatively large amplitudes, where the nonlinearity becomes
substantial. As a result, the system may have two or more coexisting
stable states of forced vibrations \cite{LL_Mechanics2004}.
Fluctuations can cause switching between these states
\cite{Dykman1979a} and thus significantly affect the overall
behavior of the system even where they are small on average. Many
features of the switching behavior and a range of phenomena and
applications related to the switching, from quantum measurements to
resonant frequency mixing and to high-frequency stochastic resonance
have been studied experimentally
\cite{Lapidus1999,Siddiqi2005a,Aldridge2005,Kim2005,Stambaugh2006,Stambaugh2006a,Siddiqi2006,Abdo2007,Almog2007}.

In this paper we analyze work fluctuations in periodically modulated
nonlinear dynamical systems coupled to a bath. We derive the
stationary state work fluctuation theorem and show that, under
fairly general assumptions, the distribution of fluctuations of work
done by the modulating force over a long time $\tau$ is Gaussian. In
common with systems close to thermal equilibrium, the work variance
$\st$ is proportional to the average work
$\langle \W\rangle$, but the proportionality coefficient is not
universal and depends on system parameters. It becomes exponentially
large in bistable systems in the range of a kinetic phase transition
where the stationary populations of the vibrational states are close
to each other. This parameter range has similarity with the region
of a first-order phase transition where molar fractions of the
coexisting phases are close to each other
\cite{Dykman1979a,Bonifacio1978,Lugiato1984}.

Large work fluctuations are related to the difference in the power
absorbed from the force in different vibrational states $i=1,2$.
Switching back and forth between the states leads to significant
fluctuations of the absorbed power. Their correlation time is
determined by the switching (transition) rates. For a small
characteristic intensity $D$ of the noise that comes from the bath,
these rates are small compared to the dynamical relaxation rate in
the absence of noise $\tr^{-1}$ and the modulation frequency
$\omega_F$. Then of primary interest are period-averaged transition
rates $\nu_{ij}$. They often display activation dependence on $D$,
with $\nu_{ij}\propto \exp(-R_{i}/D)$, where $R_{i}$ is the
characteristic activation energy of a transition $i\to j$. Since
work fluctuations accumulate power fluctuations, and the typical
accumulation time for interstate fluctuations is
$\sim\nu_{ij}^{-1}$, the exponential smallness of $\nu_{ij}$ leads
to an exponentially large factor in the variance $\st$.

We present the results for a most simple model: a nonlinear
classical dynamical system modulated by a periodic force
$F(t)=\sum\nolimits_n\tilde F(n)\exp(in\omega_Ft)$; the coupling
energy is $-F(t)q$, where $q$ is the system coordinate. The system
is additionally coupled to a bath, which leads to relaxation and
fluctuations. Work done by the force over time $\tau$ is
\begin{equation}
\label{eq:work_definition}
 \W\equiv\W(\tau)=\int\nolimits_0^{\tau}dt F(t)\dot q(t).
\end{equation}
We are interested primarily in steady-state fluctuations, i.e., we
assume that the system had come to the steady state well before the
time $t=0$ when the work (\ref{eq:work_definition}) started to be
measured. This steady state is periodic in time with modulation
period $\tau_F=2\pi/\omega_F$. We further assume that
correlations in the system decay sufficiently fast, for example,
exponentially for long times, and that the time $\tau$ largely
exceeds the characteristic correlation time of the system $t_{\rm
corr}$. Often for bistable systems $\tc\sim 1/ \nu_{ij}$.

Work fluctuations can be expressed in terms of the correlation function of
velocity fluctuations $Q(t,t')=\langle \delta \dot q(t)\delta\dot q(t')\rangle$,
where $\langle\ldots\rangle$ means ensemble average and $\delta\dot q(t)= \dot
q(t)-\langle \dot q(t)\rangle$. Because the system is in a steady periodic
state, we have $Q(t,t')=Q(t+\tau_F,t'+\tau_F)$, and therefore
\begin{equation}
\label{eq:correlation_function_general}
 Q(t,t')=\sum_nQ(n;t-t')\exp\left[in\omega_F(t+t')/2\right].
 \end{equation}

We first consider the variance of the work distribution $\st\equiv \st(\tau)=
\langle(\delta\W)^2\rangle$, where $\delta\W=\W(\tau)-\langle \W(\tau)\rangle$.
In the limit of large $\tau$
\begin{eqnarray}
\label{eq:work_variance}
 &&\st \approx
 2\pi\tau\sum_{n,m}\tilde F(n)\tilde F^*(m)\tilde
 Q\left(m-n;\frac{n+m}{2}\omega_F\right),\nonumber\\
&&\tilde
Q(n;\omega)=(2\pi)^{-1}\int\nolimits_{-\infty}^{\infty}dt\,e^{i\omega
t}Q_n(t).
 \end{eqnarray}
Here we have taken into account that the correlation functions
$Q(n;t-t')$ decay on time $t_{\rm corr}$ much smaller than $\tau$.
Therefore the limits of integration over $t-t'$ could be extended
from $-\infty$ to $\infty$.

Decay of correlations on a time scale small compared to $\tau$
allows one to simplify the expressions for higher-order moments of
$\delta\W$ in a standard way. The 3rd moment
$\langle(\delta\W)^3\rangle$ is determined by an integral over
$t_1,t_2,t_3$ of the correlator $\langle
\ddq(t_1)\ddq(t_2)\ddq(t_3)\rangle$. Because $|t_1-t_2|,
|t_1-t_3|\lesssim t_{\rm corr}$, we have $\langle(\delta\W)^3\rangle
\propto \tau$ for large $\tau$, and therefore
$\langle(\delta\W)^3\rangle/\langle(\delta\W)^2\rangle^{3/2}\propto
\tau^{-1/2}$, i.e., the 3rd moment is small for large $\tau$. The
4th moment $\langle(\delta\W)^4\rangle$ is determined by the
integral of the correlator $\langle
\ddq(t_1)\ddq(t_2)\ddq(t_3)\ddq(t_4)\rangle$. The main contribution
to this integral comes from decoupling the correlator into pairs
$\langle
\ddq(t_{n_1})\ddq(t_{n_2})\rangle\langle\ddq(t_{n_3})\ddq(t_{n_4})\rangle$
with $|t_{n_1}-t_{n_2}|, |t_{n_3}-t_{n_4}|\lesssim t_{\rm corr}$
while $|t_{n_1}-t_{n_3}|\sim\tau$ ($n_i=1,\ldots,4$). This gives
$\langle(\delta\W)^4\rangle \approx 3\langle (\delta\W)^2\rangle^2
\propto\tau^2$. Higher-order correlations in
$\langle(\delta\W)^4\rangle$ give a comparatively small contribution
$\propto \tau$. The analysis can be immediately extended to higher
moments of $\delta\W$. It shows that the overall distribution of
work fluctuations $P(\W)$ is Gaussian,
\begin{eqnarray}
\label{eq:Gaussian_work}
 P(\W)&=&(2\pi\st)^{-1/2}
 \exp\left[-(\W-\langle\W\rangle)^2/2\st\right].
 \end{eqnarray}

It follows from Eq.~(\ref{eq:work_variance}) that
$P(W)/P(-W)=\exp(2W\langle W\rangle/\st)$, as in the stationary
state fluctuation theorem for systems close to thermal equilibrium
and for modulated linear systems, and the variance of the work
distribution $\st\propto \tau\propto \langle \W\rangle$. However,
for strong periodic modulation there is no known general expression
that would relate the average velocity $\langle \dot q(t)\rangle$ to
the modulating force in terms of the correlation functions $\tilde
Q_n(\omega)$. Moreover, as we show, the ratio $\st/\langle\W\rangle$
may display sharp narrow peaks as a function of system parameters.

For weak noise, a system with two coexisting stable states $j=1,2$
mostly performs small fluctuations about these states and only
occasionally switches between them. Then to leading order in the
noise intensity $D$ the average work is a sum of partial works
$\W_{1,2}$ in each of the states weighted with the stationary
populations of the states $w_{1,2}^{\rm st}$ \cite{Dykman1979a},
\begin{eqnarray}
\label{eq:average_work}
 &&\langle\W\rangle = \sum_{j=1,2}w_j^{\rm st}\W_j,
 \qquad \W_j=\omega_F\tau\sum_n in\tilde F^*(n)
 \tilde q_j(n),\nonumber\\
 &&w_1^{\rm st}= \nu_{21}/\left( \nu_{12}+  \nu_{21}\right), \qquad
 w_2^{\rm st}=1-w_1^{\rm st}.
\end{eqnarray}
Here, $\tilde q_j(n)$ is the Fourier component of the coordinate
$q_j(t)$ in a periodic state $j$, $q_j(t)=\sum\nolimits_n\tilde
q_j(n)\exp(in\omega_Ft)$.

In contrast to the average work, the variance $\st$ has
contributions of two different types. One comes from small-amplitude
fluctuations about the stable states. It is given by the sum of
partial variances $\st_{1,2}$ weighted with the state populations.
The variances $\st_{1,2}$ can be obtained by linearizing equations
of motion about the corresponding stable state and can be written as
\begin{equation}
\label{eq:variance_intrawell}
 \st_j=C_jD\W_j, \qquad j=1,2.
\end{equation}
The constant $C_j$ depends on the details of the system dynamics.
For a system coupled to a thermal bath at temperature $T$, in the
weak-field limit we have $\st_j = 2k_BT\W_j$. However, for strong
field this relation does not hold in nonlinear systems, generally.

Fluctuations about $q_j$ become large near a saddle-node bifurcation
point where the state $j$ disappears. Here, one of the motions of
the system is slow \cite{Guckenheimer1987}, i.e., there is a ``soft
mode". Near the bifurcational (critical) position of the periodic
state $q^{(c)}_j(t)$ one can quite generally write
$q(t)-q^{(c)}_j(t)= q_{\rm sm}(t)\varkappa(t)$, where
$\varkappa(t)=\varkappa(t+\tau_F)$ is a periodic function, whereas
$q_{\rm sm}(t)$ is a slowly varying amplitude that depends on
initial conditions. After rescaling the equation of motion for
$q_{\rm sm}(t)$ can be written as \cite{Dykman2004}
\begin{eqnarray}
\label{eq:slow_variable}
 \dot q_{\rm sm} = q_{\rm sm }^2 -\eta + f(t),\qquad \langle
 f(t)f)t')=2D'\delta (t-t').
\end{eqnarray}
Here, $\eta$ is the distance to the bifurcation point, for example,
the scaled difference between the amplitude $A$ of the field and its
critical value $A_c$ at the bifurcation point. The noise $f(t)$ is
assumed to be white because of the slowness of $q_{\rm sm}(t)$; its
intensity is $D'\propto D$.

For $\eta > 0$ in the absence of noise the system
(\ref{eq:slow_variable}) has a stable state where $q_{\rm
sm}=-\eta^{1/2}$. Small fluctuations about this state have variance
$D'\eta^{-1/2}/2$ and decay over time $\tr = \eta^{-1/2}/2$.
Therefore, from Eq.~(\ref{eq:work_variance}), near the bifurcation
point where a state $j$ disappears
\begin{equation}
\label{eq:scaling_near_bifurcation}
 \st_j/W_j=\tilde C_jD/\eta.
\end{equation}
Factor $\tilde C_j$ is independent of $D$ and $\eta$, and $W_j$ does
not diverge for $\eta\to 0$.

Equation (\ref{eq:scaling_near_bifurcation}) shows that the partial
work variance scales as $\eta^{\xi}$ with the distance to the
bifurcation point. The critical exponent is $\xi=-1$.

The other contribution to $\st$ comes from fluctuation-induced
transitions between the states. The transitions lead to fluctuations
of the state populations $w_j(t)$. These fluctuations are slow,
\begin{eqnarray}
\label{eq:population_fluctuations}
 \langle\delta w_1(t)\delta w_1(t')\rangle &=&
 w_1^{\rm st}w_2^{\rm st}
 \exp\left[-\nu_{\rm tr}|t-t'|\right],\nonumber\\
 \nu_{\rm tr}&=& \nu_{12} +
 \nu_{21},
\end{eqnarray}
where $\delta w_1(t)=w_1(t)-w_1^{\rm st}=-\delta w_2(t)$. In turn,
they lead to slow fluctuations of the velocity $\dot q(t)\approx
\sum\nolimits_j\dot q_j(t)w_j(t)$ with decay time given by the
reciprocal total transition rate $\nu_{\rm tr}^{-1} \gg \tr,\tau_F$.

From Eqs.~(\ref{eq:correlation_function_general}), (\ref{eq:work_variance}),
(\ref{eq:average_work}), (\ref{eq:population_fluctuations}), the contribution to
the work variance from interstate transitions is
\begin{eqnarray}
\label{eq:variance_switching}
 \st_{\rm tr}\approx {\cal M}
 ( \nu_{\rm tr}\tau)^{-1}\left(\W_1-\W_2\right)^2,\qquad
 {\cal M}=2 w_1^{\rm st}w_2^{\rm st},
 \end{eqnarray}
and the total variance is
\begin{equation}
\label{eq:total_variance}
\st =\sum\nolimits_j w_j^{\rm st}\st_j + \st_{\rm tr}.
\end{equation}

Equation (\ref{eq:variance_switching}) is the central result of the
paper. It shows that the work variance is proportional to the
squared difference of the partial works in the stable states and is
inversely proportional to the transition rate $\nu_{\rm tr}$. The
rate $\nu_{\rm tr}\propto \exp\left(-\min\nolimits_i R_i/D\right)$
is exponentially small for small noise intensity. Respectively, the
variance (\ref{eq:variance_switching}) can be exponentially large
compared to the variance due to small fluctuations about attractors
(\ref{eq:variance_intrawell}).

Factor ${\cal M}$ in Eq.~(\ref{eq:variance_switching}) sharply
depends on the parameters of the system and the field $F(t)$. It is
small,
\begin{equation}
\label{eq:M_factor_explicit}
 {\cal M}\propto \exp\left[-|R_1-R_2|/D\right],
\end{equation}
except for a narrow range of the kinetic phase transition where the
transition activation energies are close to each other,
$|R_1-R_2|\lesssim D$. At its maximum ${\cal M}=1/2$. Factor ${\cal
M}$ determines also the intensity of super narrow peaks (of width
$\nu_{\rm tr}\ll\tr^{-1}$) in the power spectra of modulated
bistable systems and the spectra of absorption/amplification of an
additional field \cite{Dykman1979a,Dykman1989}. Its exponential
dependence on the distance to the kinetic phase transition was seen
in the simulations \cite{Dykman1990d,Dykman1994b} and experiment
\cite{Stambaugh2006a,Chan2006}.

The average work $\langle W\rangle$, Eq.~(\ref{eq:average_work}),
switches between the values $\W_1$ and $\W_2$ when the system goes
through the kinetic phase transition. Therefore the ratio
$\st/\langle\W\rangle \propto {\cal M}\nu_{\rm tr}^{-1}$ displays a
sharp peak at the transition.

The above results can be illustrated using as an example a
resonantly driven underdamped Duffing oscillator, a model that
describes a number of systems studied in recent experiments. In the absence of noise
the oscillator dynamics is described by the equation
\begin{eqnarray}
\label{eq:Duffing:general}
 \ddot q + \omega_0^2 q + \gamma q^3 + 2\Gamma \dot q=A\cos\omega_Ft.
\end{eqnarray}
We assume that the detuning of the field frequency from the
oscillator eigenfrequency $\delta\omega= \omega_F-\omega_0$ and the
friction coefficient $\Gamma$ are small: $|\delta\omega|,\Gamma \ll
\omega_0$. Then the oscillator can display bistability of forced
vibrations already for a comparatively small driving amplitude $A$,
where the vibrations remain almost sinusoidal, $q_j(t)\approx
a_j\cos(\omega_Ft+\phi_j)$ ($j=1,2$). Explicit expressions for the
amplitudes $a_{1,2}$ and phases $\phi_{1,2}$ are well known
\cite{LL_Mechanics2004}. The interstate transition rates are well
understood also
\cite{Dykman1979a,Dykman1994b,Siddiqi2005a,Aldridge2005,Stambaugh2006}.

The partial work in a stable vibrational state $j$ is
$\W_j=\Gamma\tau\omega_F^2 a_j^2$. By calculating the partial
variances $\st_j$ due to thermal noise one obtains
\begin{eqnarray}
\label{eq:partial_oscillator}
 \frac{\st_j}{\W_j}=2k_BTZ_j^{-2}\left[(Z_j-2)^2+4\Omega^2(Y_j-1)^2\right],\\
 \Omega=\delta\omega/\Gamma, \qquad
 Z_j=1+\Omega^2(Y_j-1)(3Y_j-1),\nonumber
 \end{eqnarray}
where $Y_j=3\gamma a^2_j/8\omega_F\delta\omega$ is the scaled
squared vibration amplitude. Eq.~(\ref{eq:partial_oscillator})
refers to the case where fluctuations and friction come from
coupling to the same thermal reservoir; in fact, it is not limited
to the model (\ref{eq:Duffing:general}) and applies in a general
case where the density of states of the reservoir weighted with the
interaction is smooth near $\omega_F$, cf.~\cite{Dykman1979a}.

\begin{figure}[h]
\includegraphics[width=2.8in]{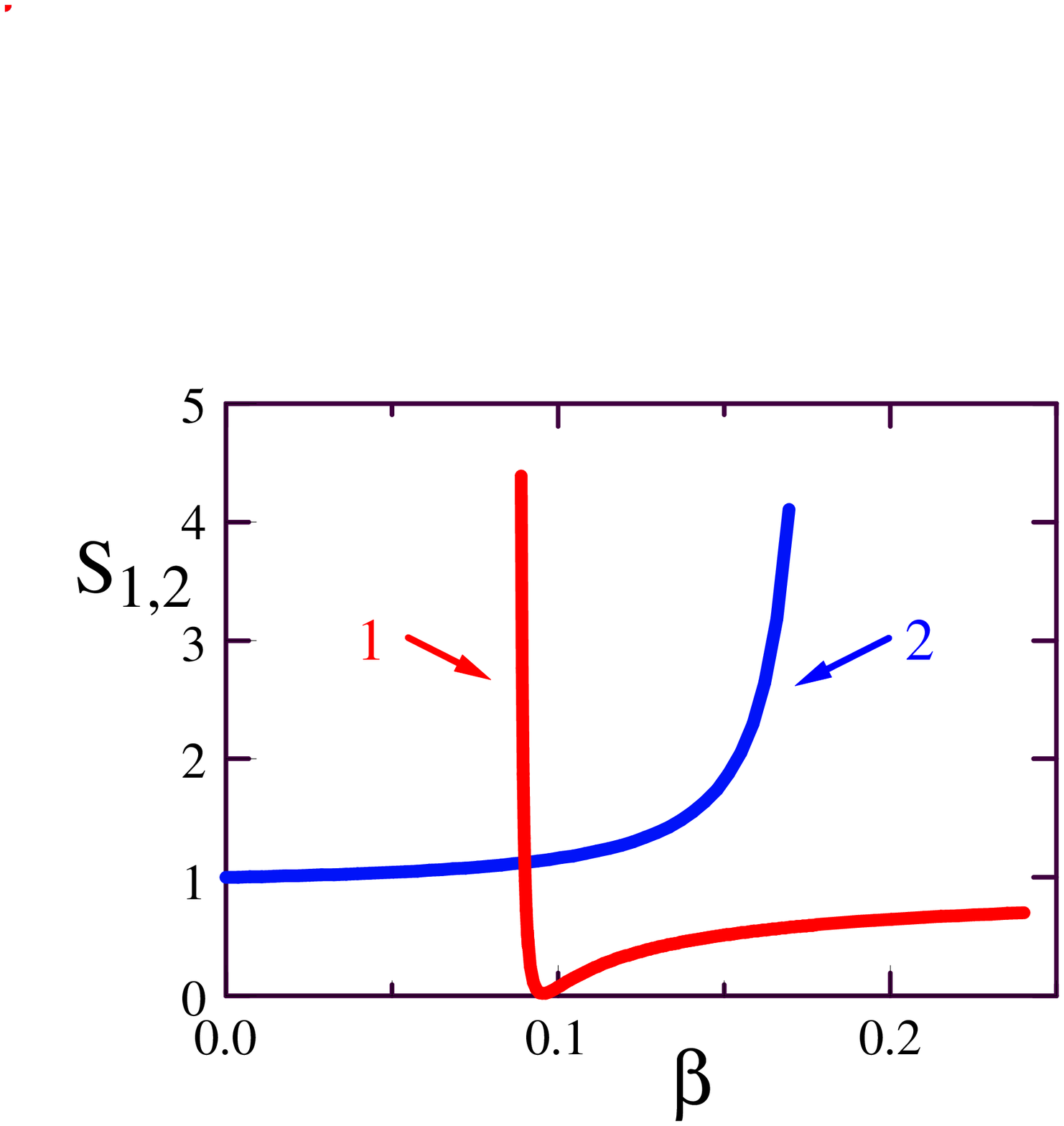}
\caption{(Color online) Scaled ratios of the partial work variance
to mean partial work $S_j=\st_j/2k_BTW_j$, $j=1,2$, as functions of
the reduced squared modulation amplitude $\beta$ for a Duffing
oscillator; $\Omega^{-1}=\Gamma/\delta\omega =0.3$. The curves 1 and
2 refer to large- and small-amplitude vibrations, respectively. The
functions $S_{1,2}$ diverge at the corresponding bifurcation points
of the oscillator, $\beta_B^{(1)}\approx 0.088$ and
$\beta_B^{(2)}\approx 0.18$. } \label{fig:oscillator_variance}
\end{figure}

It follows from Eq.~(\ref{eq:partial_oscillator}) that in the
linear-response regime, where $Y_j\propto A^2, \; Y_j\ll 1$, the
ratio $\st_j/\W_j\to 2k_BT$ is given by the standard stationary
state work fluctuation theorem. This is in agreement with the recent
theoretical and experimental results on a periodically modulated
linear system \cite{Joubaud2007a}. However, for stronger driving the
ratio (\ref{eq:partial_oscillator}) is no longer given by $2k_BT$.
When the driving amplitude $A$ or frequency $\omega_F$ approach
their bifurcational value, the ratio (\ref{eq:partial_oscillator})
diverges,
\begin{eqnarray}
\label{eq:osc_bifurcation}
 \st_j/\W_j\approx 2k_BTG_j(\Omega)\eta^{-1}, \qquad
 \eta=\beta-\beta_B^{(j)}(\Omega).
 \end{eqnarray}
Here, $\beta = 3\gamma A^2/32\omega_F^3(\delta\omega)^3$ is the
reduced field amplitude, $\beta_B^{(1,2)}$ are the bifurcational
values of $\beta$ \cite{LL_Mechanics2004}, and
$G_j=\Omega^{-2}\beta_B^{(j)}/Y_{jB}(3Y_{jB}-2)$  ($Y_{jB}$ is the
bifurcational value of the reduced squared vibration amplitude). In
agreement with Eq.~(\ref{eq:scaling_near_bifurcation}), $\st_j/\W_j$
scales as $\eta^{-1}$ with the distance $\eta$ to the bifurcation
point. The full dependence of $\st_j/\W_j$ on $\beta$ is shown in
Fig.~\ref{fig:oscillator_variance}.

It should be possible to see the scaling (\ref{eq:osc_bifurcation})
for modulated oscillators for the same conditions as the scaling of
the switching activation energies \cite{Siddiqi2005a,Stambaugh2006}.
Similarly, the exponential peak of the work variance should be seen
in experiments analogous to those where there were observed other
kinetic phase transition phenomena like super narrow peaks in the
power spectra, high-frequency stochastic resonance, and
fluctuation-enhanced frequency mixing
\cite{Stambaugh2006a,Chan2006,Almog2007}.

Experiments on the scaling and on the kinetic phase transition
should be conducted in a different way. The scaling can be observed
in the quasistationary regime. The system should be prepared in the
metastable state near a bifurcation point, and the duration of the
experiment should be shorter than the lifetime of the state. In
contrast, the exponential peak of the variance can be seen provided
the duration of the measurement exceeds the reciprocal switching
rate $\nu_{\rm tr}^{-1}$.

In conclusion, we have considered the stationary-state work
fluctuation theorem for dynamical systems modulated by a strong
periodic field. We have shown that, if the system has coexisting
states of forced vibrations, work fluctuations may become strong.
The ratio of their variance to average work is proportional to the
reciprocal rate of interstate transitions. It displays a sharp
exponentially high peak as a function of the distance to the kinetic
phase transition. In a different parameter range, near a saddle-node
bifurcation point where one of the stable vibrational states
disappears, in the quasistationary regime the variance displays
scaling with the distance to the bifurcation point.

I am grateful to E.~G.~D. Cohen for the insightful and stimulating
discussion of the fluctuation theorems. This work was supported in part by
the NSF through grant PHY-0555346.


\end{document}